\documentclass[%
 preprint, 
 amsmath,amssymb,
 aps, physrev,
]{revtex4-2}

\usepackage{graphicx}
\usepackage{dcolumn}
\usepackage{bm}
\usepackage{color}

\begin{document}

\preprint{APS/123-QED}

\title{\textbf{The candidates of 2$\bm \alpha$ condensate around the $^{\bm 16}{\bf O}$ nucleus\\
                studied by the real-time evolution method}}

\author{Y. M. Htet}
\affiliation{%
Department of Physics, Hokkaido University, Sapporo 060-0810, Japan
}%
\author{T. Oishi}
\affiliation{%
Ibaraki College in National Institute of Technology (KOSEN), Hitachinaka 312-8508, Japan
}%
\affiliation{%
RIKEN Nishina Center for Accelerator-based Science, RIKEN, Wako 351-0198, Japan
}%
\author{M. Kimura}
 \email{Contact author: masaaki.kimura@ribf.riken.jp}
\affiliation{
RIKEN Nishina Center for Accelerator-based Science, RIKEN, Wako 351-0198, Japan
}%
\author{Qing Zhao}
\affiliation{School of Science, Huzhou University, Huzhou 313000, Zhejiang, China}
\author{Bo Zhou}
\affiliation{Key Laboratory of Nuclear Physics and Ion-beam Application (MOE),
Institute of Modern Physics, Fudan University, Shanghai 200433, China}
\affiliation{Shanghai Research Center for Theoretical Nuclear Physics,
NSFC and Fudan University, Shanghai 200438, China}

\date{\today}

\begin{abstract}
\begin{description}
\item[Background] 
Searching for $\alpha$ condensation around a core nucleus, a new class of nuclear clustering, is an interesting topic. Previous theoretical studies predicted $^{16}{\rm O}+2\alpha$ condensed states. However, in those studies, the strong mixing with non-resonant states made the identification of true resonant states non-trivial.
\item[Purpose] 
To address this issue, we aim to provide a more robust theoretical verification of the $^{16}{\rm O}+2\alpha$ condensation. 
Our goal is to clearly identify the resonant states and predict their properties, such as decay widths.
\item[Method]
We employ the real-time evolution method (REM), which generates physically important basis states using the equation of motion, minimizing contamination from the continuum. The analytical continuation in the coupling constant (ACCC) was used to estimate the $\alpha$-decay widths.
\item[Results]
The present calculations show much better convergence of eigenstates, and the $0^+_3$ and $0^+_4$ states showed remarkable isoscalar monopole transition strengths, which were in good agreement with the predictions of Ichikawa et al. 
The small $\alpha$-decay widths for these states suggest that experimental observation appears feasible.
\item[Conclusion] 
The present results show the reasonable agreement with a previous work, firmly establishing the $0^+_3$ and $0^+_4$ states approximately 5 and 1 MeV below the $^{16}{\rm O}+2\alpha$ threshold as candidates for the $2\alpha$ condensation around the core nucleus $^{16}{\rm O}$. The REM proved effective in identifying these states.
\end{description}
\end{abstract}

\maketitle


\section{\label{sec:intro} Introduction}

The structure of the Hoyle state, in which three $\alpha$ particles occupy the same $0s$ orbital with a spatially extended distribution, has attracted much attention for many years~\cite{Hoyle1954,Morinaga1956,Uegaki1978,Kamimura1981,Bijker2000,Bijker2020}. 
It has been widely believed that the Hoyle state can be regarded as a Bose-Einstein condensate of $\alpha$ particles~\cite{Tohsaki2001}. 
This idea started an active search for the condensate of $\alpha$ particles~\cite{Funaki2003,Funaki2009,Schuck2016,Chernykh2007,Epelbaum2011,Epelbaum2012,Stellin2016,Lovato2016,Otsuka2022,Shen2023}, which has been extended to 4, 5 and 6$\alpha$ condensates~\cite{Wakasa2007,Funaki2008,Adachi2018,Funaki2018,Adachi2021,Zhou2023,Fujikawa2024}.

However, as schematically illustrated in the Ikeda diagram~\cite{Ikeda1968},  $n\alpha$ condensates are generally expected to appear at significantly higher excitation energies relative to the ground state of the composite system. 
This poses a substantial challenge for their experimental observation. 
As another type of $\alpha$ particle condensation, von Oertzen et al.~\cite{Oertzen2006,Koalova2006,Itagaki2007,Itagaki2010} proposed the condensation of $\alpha$ particles around an inert core nucleus. 
The presence of a core can potentially assist the formation of $\alpha$ condensates at relatively small excitation energies through attractive core-$\alpha$ interactions and the Coulomb barrier. 
This ${\rm core} + n\alpha$ condensate is a more accessible pathway for experimental exploration, although such candidates have not yet been firmly identified experimentally~\cite{Kawabata2011}.

Theoretically, Ichikawa et al.~\cite{Ichikawa2011,Ichikawa2012} have discussed the possible existence of $^{16}{\rm O}+2\alpha$ and  $^{16}{\rm O}+3\alpha$ condensates through the calculations within the $^{16}{\rm O} + n\alpha$ model space. 
They discussed that these candidates are characterized by strong isoscalar monopole transition strengths, which serve as a signature for spatially extended cluster configurations~\cite{Kawabata2007,Kanada-Enyo2007,Yamada2008}. 
They used a method where the Hamiltonian was diagonalized by randomly generated $\alpha$-particle configurations. 
While this method provided valuable insights, the randomly generated wave functions include physically unimportant configurations, leading to a strong mixing of resonances and non-resonant states. Consequently, it was unclear which of the obtained eigenstates were real resonances.

To address these limitations and further investigate the $^{16}{\rm O} $ condensate, we employ the real-time evolution method (REM)~\cite{Imai2019} to describe the $^{16}{\rm O}+2\alpha$ system. 
It is a theoretical framework that utilizes the equation of motion (EOM) of Gaussian wave packets to generate basis wave functions. 
An advantage of REM is its ability to generate physically essential basis states with less contamination from the continuum. 
Furthermore, we estimate the $\alpha$-decay width of the resonances by employing the analytical continuation in the coupling constant (ACCC)~\cite{Kukulin1977}. 
This approach provides a more robust examination of the predicted  $^{16}{\rm O}+2\alpha$ condensates in $^{24}$Mg. 

The contents of this paper are as follows:
In the next section, we explain the framework of REM applied to the $^{16}{\rm O} + 2\alpha$ system. 
The ACCC method, which was used to evaluate the $\alpha$ decay width, is also explained. 
In Section III, we provide the numerical results. 
First, we examine the convergence of the eigenstates by changing the conditions of the numerical calculations. 
Then, we discuss the properties of the resonance candidates. 
We show that the present calculations are reasonably in agreement with those of Ichikawa et al., firmly identifying the candidates for the $^{16}{\rm O} + 2\alpha$ condensation. 
The final section summarizes this work.
 
\section{\label{sec:framework}Theoretical Framework}
\subsection{Hamiltonian and model wave function}
The Hamiltonian and the model wave function used in this study are almost identical to those used in Ref.~\cite{Ichikawa2011}.
The Hamiltonian is composed of the kinetic energy, the effective two-body nuclear interaction, and the Coulomb interaction. 
The effective nuclear interaction used in this study is the Volkov No. 2 potential~\cite{Volkov1965}, with the same parameters as those employed in Ref.~\cite{Ichikawa2011}.
Its functional form is given by,
\begin{align}
V (r) = (W - MP_\sigma P_\tau ) \sum_{k=1,2} v_k \exp \left( - \frac{r^2}{c_k^2} \right), \label{eq:volkov}
\end{align}
where $P_\sigma$ and $P_\tau$ are spin and isospin exchange operators, respectively.
The Wigner parameter is set as $W = 1 - M$ with the Majorana exchange parameter $M=0.63$.
The other parameters read $v_1 = 60.650$ MeV, $v_2 = -61.140$ MeV, $c_1 = 0.980$ fm$^{-2}$, and $c_2 = 0.309$ fm$^{-2}$.

The model wave function for the $^{16}{\rm O} + 2\alpha$ system is described by the Brink-Bloch wave function~\cite{Brink1965} composed of the $^{16}$O core and two $\alpha$ clusters, 
\begin{align}
 \Phi(\bm z_1,\bm z_2) =  \mathcal{A} \{  \phi_\alpha(\bm z_1) \phi_\alpha(\bm z_2)\phi_{\rm O}(\bm z_{\rm O}) \} \label{eq:model_wf}
\end{align}
where $\mathcal{A}$ is the antisymmetrization operator which exchanges the nucleons belonging to different clusters. 
Each $\alpha$ cluster wave function $\phi_{\alpha}(\bm z)$ describes four nucleons (two protons and neutrons with spin up and down) in a $(0s)^4$ configuration. 
\begin{align}
\phi_{\alpha}(\bm z) = \mathcal{A}\{ \varphi(r_1, \bm z)\chi_{p\uparrow} \cdots \varphi(r_4, \bm z)\chi_{n\downarrow} \}, 
\end{align}
where $\chi_{p\uparrow}$, etc., denote the spin-isospin states of the nucleons, and the single-particle Gaussian wave packet $\varphi(r, \bm z)$ is defined as:
\begin{align}
\varphi(r, \bm z) = \left(\frac{2\nu}{\pi}\right)^{3/4} \exp\left\{ -\nu \left( r - \frac{\bm z}{\sqrt{\nu}} \right)^2 + \frac{1}{2} z^2 \right\}.
\end{align}
Here, $\nu$ is the size parameter of the $\alpha$ particle, and is chosen to reproduce the observed radius of $^4$He ($\nu = 0.275$fm$^{-2}$). 
The complex-valued three-dimensional vector $\bm z$ represents the center of the Gaussian wave packet in phase space, i.e., the real/imaginary part of $\bm z$ corresponds to the spatial position/momentum of an $\alpha$ particle.
Note that the only difference between this wave function and Ref.~\cite{Ichikawa2011}'s lies in the Gaussian centroid, which is parametrized by real-valued or complex-valued three-dimensional vectors.

The $^{16}$O cluster is approximated as a $4\alpha$ system,
\begin{align}
\phi_{\rm O}(\bm z_{\rm O}) &= \mathcal{A}\{ \phi_{\alpha}(\bm z_3) \cdots \phi_{\alpha}(\bm z_6) \}, \\
\bm z_{\rm O} &= \frac{1}{4}\left(\bm z_3+\cdots +\bm z_6\right),
\end{align}
where the centroids of four $\alpha$ particles ($\bm z_3,...,\bm z_6$) are fixed at the vertex of a tetrahedron with a side length of 0.5 fm, 
and only their center-of-mass ($\bm z_{\rm O}$) is treated as a free parameter.
Furthermore, we impose the condition that the center of mass of the system is located at the origin, i.e., $\bm z_1+\bm z_2+4\bm z_{\rm O} = 0$.
Thus, the model wave function Eq.~\eqref{eq:model_wf} has two independent parameters, $\bm z_1$ and $\bm z_2$.

\subsection{real-time evolution method}
The most significant feature of this work is that the basis functions are generated by using EOM, while they were generated stochastically (randomly) in Ref.~\cite{Ichikawa2011}. 
We regard $\bm z_1$, $\bm z_2$, and $\bm z_o$ as the time-dependent parameters, and derive the EOM from the time-dependent variational principle: 
\begin{align}
i\hbar \sum_{j=1}^2 \sum_{\sigma=x,y,z} C_{i\rho; j\sigma} \frac{d z_{j\sigma}}{dt} = \frac{\partial E_{int}}{\partial  z^*_{i\rho}},
\end{align} 
where
\begin{align}
E_{int} := \frac{\langle \Phi(\bm z_1,\bm z_2) | \hat{H} | \Phi(\bm z_1,\bm z_2)\rangle}
               {\langle \Phi(\bm z_1,\bm z_2) | \Phi(\bm z_1,\bm z_2)\rangle},\\
C_{i\rho; j\sigma} := \frac{\partial^2 \ln \langle \Phi (\bm z_1,\bm z_2)| \Phi (\bm z_1,\bm z_2)\rangle}
                        {\partial z^*_{i\rho} \partial z_{j\sigma}}.
\end{align}
Starting from an arbitrary initial wave function at $t = 0$, we solve this EOM to calculate the time evolution of $\bm z_1$ and $\bm z_2$.
As a result, the EOM produces the time-dependent vectors $\bm z_1(t)$ and $\bm z_2(t)$. 
We expect that if time evolves long enough, the ensemble of wave functions $\Phi(\bm z_1(t),\bm z_2(t))$ spans a good model space to describe both bound states and resonances. Indeed, this method successfully described many different cluster states~\cite{Imai2019,Zhou2020,Shin2021,Zhao2022,Zhao2022-2,Yang2023}
Hence, the bound and resonant states for $^{16}{\rm O}+2\alpha$ system may be reasonably described by their superposition:
\begin{align}
\Psi = \int^T_0 dt \sum_{K=-J}^J f_K(t)P^{J\pi}_{MK} \Phi(t), \label{eq:rem_wf1}
\end{align}
where $T$ is the total time-propagation length, $P^{J\pi}_{MK}$ is the parity and angular momentum projection operator, and $\Phi(t)$ is a shorthand of $\Phi(\bm z_1(t),\bm z_2(t))$. 
In the practical calculations, we discretize time as,
\begin{align}
\Psi = \sum_i \sum_{K=-J}^J f_{iK}P^{J\pi}_{MK} \Phi(t_i). \label{eq:rem_wf2}
\end{align}
The amplitude $f_{iK}$ and eigen-energy are calculated by the Hill-Wheeler (HW) equation~\cite{Hill1953,Griffin1957},
\begin{align}
    \sum_{jK'}(H_{iK;jK'} -EN_{iK;jK'})f_{jK'}   = 0,
\end{align}
where
\begin{align}
    H_{iK;jK'} &:= \langle P^{J\pi}_{MK} \Phi(t_i)|H|P^{J\pi}_{MK'} \Phi(t_j)\rangle,\\
    N_{iK;jK'} &:= \langle P^{J\pi}_{MK} \Phi(t_i)|P^{J\pi}_{MK'} \Phi(t_j)\rangle.
\end{align}

\subsection{Calculation procedure}
Here, we summarize the practical calculation procedure adopted in this work.

\begin{enumerate}
    \item \textbf{Imaginary-time evolution:} We evolve a randomly generated wave function in imaginary time,
    and obtain a state excited by $E^*:=E_{int}-E_{min}$ (MeV), which serves as the initial state for the real-time evolution. 
    Here, $E_{min}$ denotes the lowest intrinsic energy, which is also calculated by the imaginary time evolution.
    We tested several choices of $E^*$ from 20 to 50 MeV.

    \item \textbf{Real-time evolution:} We calculate the real-time evolution of the system according to the EOM. 
    To prevent $\alpha$ particles from escaping to infinity, a spherical wall that reflects the $\alpha$ particles is placed at a distance $R$ from the origin of the coordinate. 
    Several values of $R$ (from 4 fm to 9 fm) were tested. 
    The time evolution was calculated with a discrete time step of $\Delta t = 0.002$ fm/$c$, and the wave function is sampled every 2000 steps (4 fm/$c$) up to a maximum evolution time of $T=4000$ fm/$c$, resulting in a maximum of 1000 sampled wave functions.

    \item \textbf{Selection of basis functions:} To avoid redundancy in the basis set, sampled wave functions with an overlap greater than 0.95 are removed. 
    Namely, for any combination of $i$ and $j$, the following condition must be satisfied,
    $$\frac{|\langle P^{0^+}\Phi(t_i)|P^{0^+}\Phi(t_j)\rangle|^2}
    {\langle P^{0^+}\Phi(t_i)|P^{0^+}\Phi(t_i)\rangle\langle P^{0^+}\Phi(t_j)|P^{0^+}\Phi(t_j)\rangle}<0.95,$$
    where $P^{0^+}$ denotes the projection operator to $J^{\pi}=0^+$ state.
    The number of remaining basis functions typically ranges from 300 to 400, depending on the calculation conditions.

    \item \textbf{Solution of HW equation:} The selected basis wave functions are projected to the $J^\pi=0^+$ state. 
    The eigenenergies and eigenstates are obtained by solving the HW equation. 
    The calculations are repeated with different values of the rebound radius $R$ and the initial excitation energy $E^*$ to check the convergence of the results.
\end{enumerate}

\subsection{Analytical Continuation in the Coupling Constant (ACCC)}
The ACCC~\cite{Kukulin1977} is a practical method to evaluate the resonance energy and width. 
It introduces an attractive auxiliary potential to artificially bind the resonance states.
As an auxiliary potential, we vary the strength of the exchange term of the Volkov potential (Eq.~\eqref{eq:volkov}) as,
\begin{align}
M = 0.63 - \mu, \quad W=1-M = 0.37 + \mu,
\end{align}
where $\mu$ is a positive number. Note that this does not change the binding energy of the $\alpha$ particle but makes the $\alpha$-$\alpha$ 
 and $^{16}{\rm O}$-$\alpha$ interactions more attractive. 
Consequently, a larger value of $\mu$ gives deeper binding of $^{16}{\rm O}$, $^{16}{\rm O}+\alpha$ and $^{16}{\rm O}+2\alpha$ systems.

Let us denote the energy of the ground state of the $^{16}{\rm O}+\alpha$ system obtained with a certain value of $\mu$ as $E_{\rm O+\alpha}(\mu)$, and the energy of an excited state of the $^{16}{\rm O}+2\alpha$ system as $E_{\rm O+2\alpha}(\mu)$.
Hence, $E(\mu) := E_{\rm O+2\alpha}(\mu) - E_{\rm O+\alpha}(\mu)$ is the $\alpha$ decay $Q$-value.
Then, consider an excited state of the $^{16}{\rm O}+2\alpha$ system that is unbound at the physical point ($\mu = 0$), i.e.,  $E(0) > 0$. 
As the value of $\mu$ increases, this state becomes bound. That is, $E(\mu_0) =0$ for a certain value $\mu_0$, and $E(\mu) < 0$ holds for $\mu > \mu_0$.
Note that in the region $E(\mu) < 0$ ($\mu > \mu_0$), the eigenstate can be accurately calculated within a finite spatial domain. 
However, this is not the case for $\mu < \mu_0$, where the exact solution is not square integrable.
Therefore, the ACCC method analytically continues the solution from the region $\mu > \mu_0$ into the region $\mu < \mu_0$, allowing us to determine the resonance energy and width at the physical point.

Practically, following the procedure in Ref.~\cite{Tanaka1997,Takatsu2023}, we introduce a variable $x = \sqrt{\mu - \mu_0}$. Then, $\mathcal{E}(x):=E(\mu)$ is written as,
\begin{align}
\mathcal{E}(x) &= \frac{\hbar^2}{2m}k^2(x),\label{eq:accc1}\\
k(x) &= i \frac{c_1 x + \dots + c_M x^M}{1 + d_1 x + \dots + d_N x^N}, \label{eq:accc2}
\end{align}
where $m$ is the reduced mass, and the wave number $k(x)$ is expressed by the rank $[M/N]$ Pad\`e approximation. 
The coefficients $c_1,...,c_M,d_1,...,d_N$ are determined by the fitting in the region $\mu > \mu_0$. 
Finally, Eqs.~\eqref{eq:accc1} and \eqref{eq:accc2} are extrapolated to the physical point, $\mu =0$. 
That is, the resonance energy $E_R$ and width $\Gamma_R$ are evaluated at the physical point $x_0 = i\sqrt{\mu_0}$,
\begin{align}
\mathcal{E}(x_0)=\frac{\hbar^2 k^2(x_0)}{2\mu} = E_R - i \frac{\Gamma_R}{2},
\end{align}
where $E_R$ and $\Gamma_R$ denote the resonance energy and width, respectively.
In the practical calculation, we found that $M=N=6$ is large enough to obtain the stable results.

\section{Results and Discussions}
\subsection{Convergence of the eigenstates}
Here, we explain how the calculation results converge with a long-time evolution. 
We also show how the results depend on the numerical conditions, the rebound distance $R$ and the intrinsic energy $E^*$.

\begin{figure}[h]
\includegraphics[width=0.5\hsize]{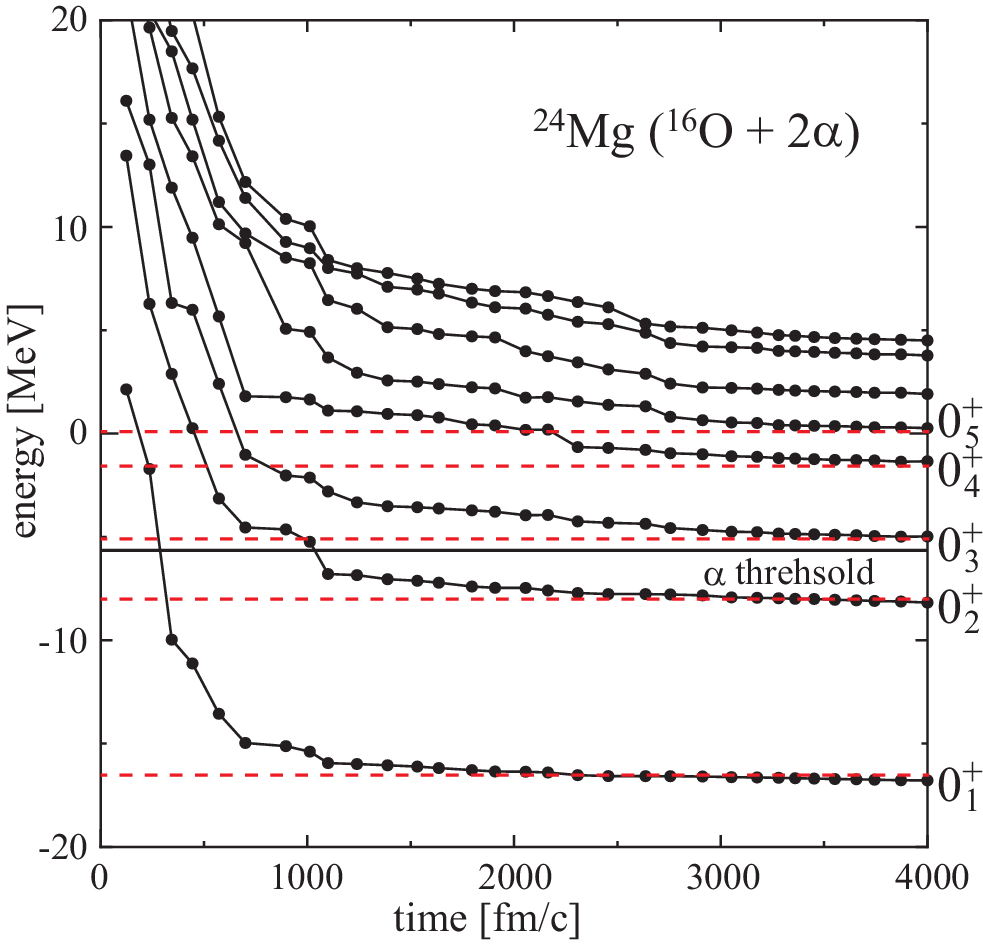}
\caption{\label{fig:time-energy} The energies of the $0^+$ states obtained with the time evolution up to $T$ [fm/$c]$.
The dotted line shows the calculated $\alpha$ decay threshold. 
The red dashed lines show the energies of the bound states and resonance candidates suggested in Ref.~\cite{Ichikawa2011}.}
\end{figure}

Figure~\ref{fig:time-energy} shows how the eigenenergies converge as a function of the evolution time $T$, under the condition of $E^*=35$ MeV and $R=7$ fm. 
The $0^+_1$ and $0^+_2$ states are the bound states lower than the $\alpha$-decay threshold.
After long-time evolution, their energies are slightly lower than those calculated by Ichikawa et al.~\cite{Ichikawa2011},
which indicates an improvement by REM compared to the random generation of the basis wave function.

For unbound states above the threshold (the $0^+_3$ and higher states), the energy slowly decreases over time due to coupling with the continuum.
Nevertheless, the obtained result plausibly agrees with the candidates of $^{16}{\rm O}+2\alpha$ resonances proposed in Ref.~\cite{Ichikawa2011}. 
Compared to the random basis generation, the coupling with the continuum is significantly weaker in REM. 
In Ref.~\cite{Ichikawa2011}, the random generation yielded many $0^+$ states below the $^{16}{\rm O} + 2\alpha$ threshold, but most of which were discarded as non-resonant solutions due to their instability. 
The REM results are free from such instability, allowing for the unique identification of resonance states.
For highly excited states above the $^{16}{\rm O} + 2\alpha$ threshold, however, the coupling with non-resonant states is strong, and we could not clearly distinguish between resonant and non-resonant states.

\begin{figure}[h]
\includegraphics[width=0.5\hsize]{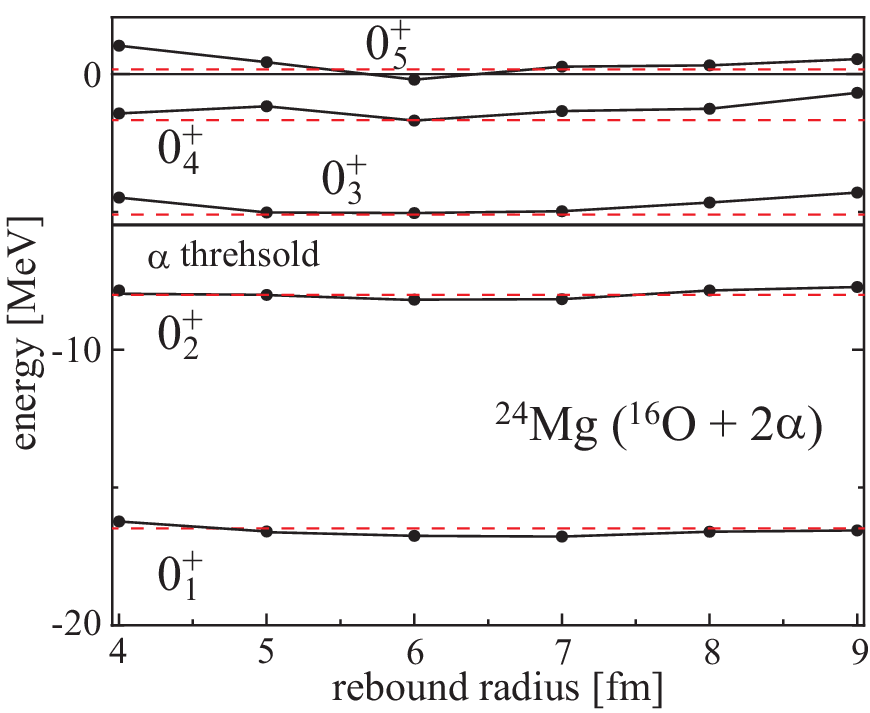}
\caption{\label{fig:radi-energy} The energies of the five lowest $0^+$ states obtained after the time evolution of $T=4000$ [fm/$c$]
with the same energy $E^*=35$ MeV, but different rebound radii $R=4,...,9$ fm.
The dotted line shows the calculated $\alpha$ decay threshold. 
The red dashed lines show the energies of the bound states and resonance candidates suggested in Ref.~\cite{Ichikawa2011}.}
\end{figure}

Next, we examine the convergence by varying the rebound radius $R$. 
Fig.~\ref{fig:radi-energy} shows how the eigenenergies depend on the reflection radius $R$ with fixed $E^*=35$ MeV and $T=4000$ fm.
It was found that the energies of the first to fifth $0^+$ states do not change significantly depending on the rebound radius. 
These suggest that they are resonant states (the $0^+_1$ and $0^+_2$ states are bound).
Looking more closely at the $0^+_1$ and $0^+_2$ states, they are most deeply bound when the rebound radius is $R=6$ or $7$ fm. 
At smaller rebound radii, $R=4$ or $5$ fm, the spatial size seems to be too small, whereas at larger radii of $R=8$ or $9$ fm, a sufficient number of states could not be generated within the finite time evolution ($4000$ fm/c).

\begin{figure}[h]
\includegraphics[width=0.5\hsize]{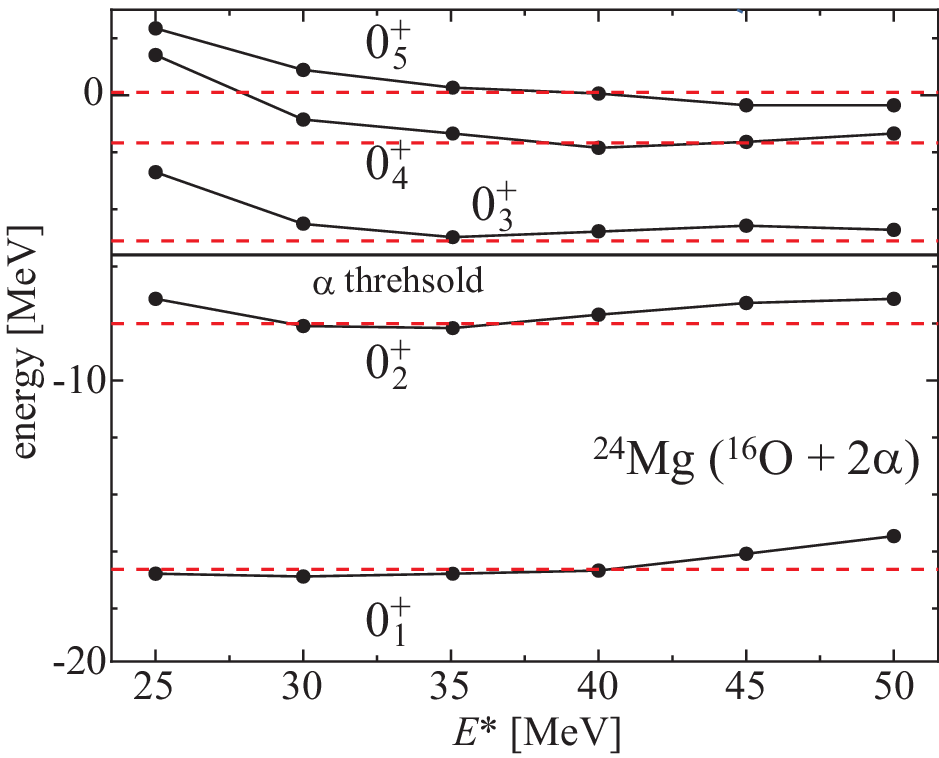}
\caption{\label{fig:eng-energy} The energy of the five lowest $0^+$ states obtained after the time evolution of $4000$ [fm/$c$]
with the same rebound radius $R=7$ fm, but with different intrinsic energies $E^*=25,...,50$ MeV were applied for solving EOM .}
\end{figure}

Fig. 3 shows how the eigenenergies depend on the intrinsic excitation energy $E^*$ applied in solving the EOM.
First, we see that approximately consistent and reliable results are obtained in the region of $E^*$ from 30 to 40 MeV. 
At $E^* = 25$ MeV, the energy of the excited state is overestimated. 
Since $E^*$ is too small, the $\alpha$ particle could not cross the Coulomb barrier and moved only within a space smaller than $R = 7$ fm. 
On the other hand, for $E^* \ge 45$ MeV, the energies of the low-lying excited states are overestimated, which is likely because the generated ensemble did not span the model space large enough to describe the low-lying states due to a higher value of $E^*$.
Thus, the results obtained at $R = 7$ fm seem reasonable, and the discussion in the next section is based on them.

\subsection{Properties of the resonance candidates}
Here, we examine the properties of the resonance candidates obtained in the previous section. 
Table~\ref{tab:table1} summarizes their properties: the energies, widths, isoscalar monopole transition strength, and radii.

\begin{table}[h]
\caption{\label{tab:table1}%
The properties of the first five $0^+$ states: 
energy (MeV), width (MeV), isoscalar monopole transition strength from the ground state (fm$^4$), and radius (fm).
The data for the $0^+_3$ and $0^+_4$ states are evaluated by ACCC, while ACCC was not applicable for the $0^+_5$ state.}
\begin{ruledtabular}
\begin{tabular}{cccccc}
state & $E$ & $\Gamma$ & $B(IS0)$ &$\sqrt{\langle r^2\rangle}$ & $\sqrt{\langle r^2\rangle_{\rm val}}$ \\
\colrule
$0^+_1$ & -16.78& --           & --    & 2.67 &  3.47\\ 
$0^+_2$ & -8.16 & --           & 28.4  & 2.70 &  3.53\\
$0^+_3$ & -4.95 & $\leq$ 0.01  & 145.0 & 2.97 &  4.13\\
$0^+_4$ & -1.26 & 0.17         & 160.7 & 2.94 &  4.08\\
$0^+_5$ &  0.37 & --           & 4.28  & 3.03 &  4.28\\
\end{tabular}
\end{ruledtabular}
\end{table}

For the bound states ($0^+_1$ and $0^+_2$ states), as mentioned above, the present calculations yield slightly deeper binding energies compared to Ref.~\cite{Ichikawa2011} ($-16.63$ and $-8.01$ MeV, respectively). 
In contrast, the energies of the $0^+_3$ and $0^+_4$ states are slightly higher. 
This is due to the decoupling of the continuum states by the ACCC method. 
Their widths, especially that of the $0^+_3$ state, are small, suggesting that they are experimentally observable. 
However, we must consider the proton decay channel, whose threshold is located at $-5.09$ MeV. 
Although the proton decay is outside our model space, its impact on the $0^+_3$ state seems to be minor,
since the decay $Q$-value is quite small. 
In contrast, its influence on the $0^+_4$ state may be non-negligible, and evaluating it with a microscopic model is a future task.
For the $0^+_5$ state, we were unable to obtain the stable solution of ACCC because its energy is too high above the $\alpha$ threshold to be extrapolated from the bound state region.

Next, we examine the isoscalar monopole transition strength, $B(IS0)$, which characterizes spatially extended cluster states~\cite{Yamada2008}. 
As pointed out by Ichikawa et al., the transition strengths of the $0^+_3$ and $0^+_4$ states are remarkable. 
Quantitatively, the present results are very close to their results (143.80 and 152.98 fm$^4$, respectively), 
which further supports the argument that they are candidates for $^{16}{\rm O}+2\alpha$ condensate.

Finally, we discuss the radii. 
Although the radius cannot be strictly defined for resonance states, that calculated within the bound-state approximation provides a measure of the spatial extent of the interacting clusters. 
While the radii of the bound states are around 2.7 fm, the radii of the $0^+_3$ and $0^+_4$ states are approximately 3.0 fm, 
showing some increase. This is also in good agreement with the values obtained by Ichikawa et al.
The radii of the $0^+_3$  and $0^+_4$ states seem small compared to that of the Hoyle state of $^{12}$C (3.7 fm).
However, this comparison may not be straightforward since the contribution from the inert core, $^{16}$O, should be subtracted. 
Therefore, we define the distribution radius of the valence $2\alpha$ particles around the core as follows:
\begin{align}
\langle r^2 \rangle_{val} := \frac{1}{8}\sum_{i\in \rm valence}r_i^2 = 
3\langle r^2 \rangle_{\rm O+2\alpha} - 2\langle r^2 \rangle_{\rm O},
\end{align}
where $\langle r^2\rangle_{\rm O+2\alpha}$ and $\langle r^2\rangle_{\rm O}$ denote the square radii of 
the $^{16}{\rm O}+2\alpha$ system and the $^{16}{\rm O}$ core, respectively.
Based on this, the distribution radii of the $2\alpha$ particles around the core are calculated in Table~\ref{tab:table1}. 
It can be seen that the distribution radii of the valence $\alpha$ particles in the $0^+_3$ and $0^+_4$ states are even larger than that of the Hoyle state. 
Furthermore, they (4 fm) roughly coincide with the expected radius of the Coulomb barrier of the $^{16}{\rm O}+\alpha$, 
suggesting the $0^+_3$ and $0^+_4$ states have $\alpha$ particles trapped between the surface of the core nucleus and the Coulomb barrier.

\section{Summary}
The search for ${\rm core} + n\alpha$ condensates where $\alpha$ particles condense around an inert core nucleus, is an interesting problem, offering a novel type of $\alpha$ clustering. 
Previous theoretical works suffered from significant mixing with physically unimportant configurations, making the clear identification of real resonances difficult.

To address these limitations and further investigate the predictions of the $^{16}$O+2$\alpha$ condensate by Ichikawa et al., we employed REM to generate physically important basis wave functions with less contamination from the continuum. 
Furthermore, we applied the ACCC to estimate the $\alpha$-decay widths of the resonance candidates.

The present numerical results show reasonable agreement with those of Ichikawa et al. and firmly establish the candidates for $^{16}{\rm O}+2\alpha$ condensation. 
The REM successfully generated cleaner ensembles, aiding in the identification of resonance states. 
Especially, the $0^+_3$ and $0^+_4$ states were identified as strong candidates. 
These states exhibited remarkably enhanced isoscalar monopole transitions from the ground state, quantitatively close to previous findings, strongly supporting their nature as spatially extended $^{16}{\rm O}+2\alpha$ condensate. 
The valence 2$\alpha$ distribution radii for these states (around 4 fm) were comparable with the expected radius of the Coulomb barrier around $^{16}$O, suggesting the $\alpha$ particles might be trapped by the barrier. 
The calculated $\alpha$-decay widths of the $0^+_3$ and $0^+_4$ states were small, suggesting that they should be experimentally observable.

\begin{acknowledgments} 
This work was supported by JSPS KAKENHI Grant Number 22H01214 and by the RIKEN TRIP initiative (Nuclear Transmutation). 
\end{acknowledgments}

\nocite{*}

\bibliography{export}

\end{document}